\documentclass[a4paper,11pt]{article}
\usepackage{jcappub}
\usepackage{epsfig}
\usepackage{graphicx}
\usepackage{amssymb}
\usepackage[tight]{subfigure}

\arxivnumber{2402.04743}

\title{\boldmath Temperature evolution in the Early Universe and  freeze-in at stronger coupling}

\author[a]{Catarina Cosme, }
\author[b]{Francesco Costa, }
\author[c]{and Oleg Lebedev}

\affiliation[a]{Univ Coimbra, Faculdade de Ci\^encias e Tecnologia da Universidade de Coimbra \\
  and CFisUC, Rua Larga, 3004-516 Coimbra, Portugal}

\affiliation[b]{Institute for Theoretical Physics, Georg-August University G\"ottingen, \\
  Friedrich-Hund-Platz 1, G\"ottingen D-37077, Germany  }

\affiliation[c]{Department of Physics and Helsinki Institute of Physics,\\
  Gustaf H\"allstr\"omin katu 2a, FI-00014 Helsinki, Finland}
  
\emailAdd{ccosme@uc.pt}
\emailAdd{francesco.costa@uni-goettingen.de}
\emailAdd{oleg.lebedev@helsinki.fi}

\abstract{ Dark matter freeze-in at stronger coupling is operative when  the Standard Model (SM) bath temperature never exceeds the dark matter mass. An attractive feature of this scenario is that it can be probed by direct detection
 experiments as well as at the LHC.
 In this work, we show how the mechanism can be realized in a simple UV complete framework, emphasizing the role of the maximal temperature of the SM thermal bath. We demonstrate that the maximal temperature can coincide with the reheating temperature or be close to it such that dark matter production is always Boltzmann--suppressed. This possibility is realized, for example, if the inflaton decays primarily into feebly interacting right-handed neutrinos, which subsequently generate the SM thermal bath.  In this case, the SM sector temperature remains constant over cosmological times prior to reheating.
}

\begin{document}
 
\maketitle
\flushbottom

 \section{Introduction}

 The dark matter (DM) puzzle remains one of the outstanding questions in modern physics.
 While thermal dark matter in the form of a weakly-interacting-massive-particle (WIMP) emerges naturally in many extensions of the Standard Model, the lack of observational evidence in support of WIMPs \cite{LZ:2022ufs}
 motivates us to explore further options. 
 In particular, dark matter can be non-thermal and produced in a variety of ways. One popular mechanism is known as ``freeze-in'', whereby DM is gradually produced by the SM thermal bath
 \cite{Dodelson:1993je,McDonald:2001vt,Kusenko:2006rh,Hall:2009bx}.
 Its conventional version requires a feeble coupling between the SM sector and dark matter, as well as a high enough reheating temperature.
  This scenario assumes a zero DM abundance before the thermal production starts, which appears quite unlikely in the world with gravity \cite{Lebedev:2022cic}. The freeze-in mechanism is also notoriously difficult to test, if possible at all.

 The recently proposed ``freeze-in at stronger coupling'' \cite{Cosme:2023xpa} addresses some of the problematic aspects of traditional freeze-in models. It requires a low reheating temperature such that DM production 
 is Boltzmann-suppressed. In this case, the coupling between DM and the SM sector  is allowed to be up to ${\cal O}(1)$ without affecting  the non-thermal nature of dark matter. This makes DM potentially observable in direct detection experiments as well as 
 at the LHC. 
 
 The low reheating temperature 
  allows for dilution of 
 gravitationally produced dark matter, which otherwise is problematic in most models \cite{Lebedev:2022cic}. Specifically, all particles are produced by  gravity or via gravity-induced interactions both during and after inflation.
 This is particularly problematic for scalar DM, whose field fluctuations build up during inflation \cite{Starobinsky:1994bd} resulting in DM overproduction, unless a suppression or  dilution mechanism is invoked. 
 Even if DM production gets suppressed during inflation, e.g. via a large Hubble-induced dark matter mass, it gets produced during preheating all the more efficiently.  Also, classical and quantum gravitational effects induce higher dimensional operators, which couple dark matter  $s$ to the inflaton $\phi$ \cite{Lebedev:2022ljz}, e.g.
 \begin{equation}
C \; {\phi^4 s^2 \over M_{\rm Pl}^2} \;,
\end{equation}  
where $C$ is a Wilson coefficient. During the inflaton oscillation epoch, this operator produces dark matter very efficiently in traditional (large field) inflation models, normally leading to a totally dark Universe.

The resulting abundance can be made consistent with observations only if the Universe undergoes a very long period of inflaton matter-domination, which dilutes all relativistic relics,
or if the Wilson coefficient happens to be  tiny, which cannot be guaranteed in the absence 
of  a UV complete quantum gravity framework.
The constraint can be put in the form  \cite{Lebedev:2022cic}
\begin{equation}
 \Delta_{\rm NR} \gtrsim 10^6\; C^2  \;{       \phi_0^8  \over    H_{\rm end}^{5/2}  \, M_{\rm Pl}^{11/2}} \; {m_s \over {\rm GeV}} \;,
  \label{c4-bound}
 \end{equation}
where $\Delta_{\rm NR} \equiv
\left( {H_{\rm end}\over H_{\rm reh}}\right)^{1/2}  $ characterizes the duration of the matter-dominated expansion period. Here $H_{\rm end} $ and $  H_{\rm reh}$ are the Hubble rates at the end of inflation and at reheating, respectively; 
$\phi_0$ is the inflaton field value at the end of inflation and $m_s$ is the DM mass. 
Taking $\phi_0 \sim M_{\rm Pl}$ and $H_{\rm end} \sim 10^{14}$ GeV, one obtains a very strong bound
$
 \Delta_{\rm NR} \gtrsim 10^{17 }\; C^2 \;  {m_s \over {\rm GeV}} \;,
$
 which implies a low reheating temperature $\Delta_{\rm NR} \gg 1 $  (unless dark matter is super-light, $m_s \ll \;$eV,  or $C^2$ happens to be  vanishingly small).
 Similar considerations apply to fermionic dark matter, although the  constraints are weaker \cite{Koutroulis:2023fgp}.

Freeze-in at stronger coupling operates in models with low reheating temperature $T_R$,  thereby avoiding the problem of  gravitational DM overproduction. The dark relics generated during and immediately after inflation 
are diluted away such that one may assume their negligible abundance at the onset of thermal production. If dark matter is heavier than $T_R$, its production is Boltzmann-suppressed and 
its coupling to the SM fields can be significant. 
On the other hand, the DM abundance becomes  very sensitive to the thermal history of the Universe. 
In particular, it is produced most efficiently when the SM bath temperature reaches its maximum, which generally happens before reheating.

In our previous work \cite{Cosme:2023xpa}, we have resorted to the low energy effective description and, in particular, to  the instant reheating approximation, which assumes that dark matter is mostly produced at $T_R$ and the preexisting abundance can be neglected.
The  goal of the current work is to study under what circumstances this assumption is adequate and, more generally, explore 
cosmological frameworks where freeze-in at stronger coupling can be realized. We find broad classes of models in which
the SM bath temperature stays constant over cosmological times such that  
 the maximal and reheating temperature are naturally close.  These allow for a reliable calculation of freeze-in DM abundance, without the problem of initial conditions inherent in high $T_R$ models,
 and justify our previous results.

 \section{Evolution of the Standard Model sector temperature}

 Little is known as to how exactly the SM fields were produced after inflation. 
 They could be generated through their direct couplings to the inflaton \cite{Kolb:1990vq} or result from interactions with  a secondary field, e.g. decay thereof.
 The SM sector temperature exhibits very different evolution in these cases.
 
 Let us consider a general case of the SM particle production from decay of a field $\chi$, which does not necessarily dominate the energy density of the Universe. 
 The SM quanta are typically produced in the relativistic regime, so we will assume that the SM energy density  $\rho$ scales as radiation. 
 The energy density of the $\chi$ field is $\rho_\chi$, whose scaling is   $a^{-n}$, while that of the Hubble rate $H$ is $a^{-m}$, with $a$ being the scale factor.
 The values of $n$ and $m$ are, in general, unrelated and depend on further details. Denoting the decay width of $\chi$ by 
 $\Gamma_\chi$, we arrive at the following system:
\begin{eqnarray}
&& \dot \rho + 4 H\rho= \Gamma_\chi \rho_\chi \;,  \nonumber \\
&& H= H_0 /a^m \;, \nonumber \\
&& \rho_\chi = \rho_\chi^0 / a^n \;, \label{system}
\end{eqnarray}
The label  $0$ refers to the initial moment  $a_0=1$  corresponding to the end of inflation and $\rho (1)=0$. 
We assume  here that the SM sector does not contribute significantly to the energy balance of the Universe and that $n,m$ are constant in the regime of interest.

The solution for $\rho(a)$ is easily found as follows: $\dot \rho + 4 H\rho = {1\over a^4} \, { d\over dt} (a^4 \rho)$ and $dt = da/(aH)$, so that
\begin{equation}
a^4 \rho =  \Gamma_\chi \rho_\chi^0\; \int {da \over aH} \,a^{4-n} \;,
\end{equation}
and our boundary condition $\rho (1)=0$ requires
\begin{equation}
  \rho(a) =  {\Gamma_\chi \rho_\chi^0 \over  (4-n+m)H_0}\; \left[    {1\over a^{n-m}} -{1\over a^4}  \right] ~~\rightarrow ~~{\Gamma_\chi \rho_\chi^0 \over  (4-n+m)H_0 } \; {1\over a^{n-m}} 
  \label{rhoSM}
\end{equation}
at $a\gg 1$ since $n-m <4$ for all cases of interest. In practice, this approximation becomes sensible already at $a= {\cal O}(1)$.

The above scaling is different from the usual $1/a^4$ law  applicable when the SM radiation dominates.
In fact, $n-m$ can be of either sign or zero. The SM energy density can thus grow in time or stay constant before reheating.
Since thermalization in the SM sector is fast, one can associate the SM thermal bath temperature with $\rho^{1/4}$, i.e.
\begin{equation}
T\simeq       \left({30\over g_* \pi^2}\right)^{1/4}   \rho^{1/4} \;,
\end{equation} 
where $g_* $ is the number of the SM degrees of freedom.
Therefore, the SM sector temperature is allowed to grow in time or stay constant before reheating.  The latter is defined as the time when the SM energy density starts dominating the energy balance of the Universe and is associated with temperature $T_R$.

If the SM fields are produced directly by the inflaton, $n=2m$ and $\rho \propto a^{-m}$. This corresponds to the $a^{-3/2}$ and $a^{-2}$ scaling for the $\phi^2$ and $\phi^4$
inflaton potentials, respectively. In this case, the SM temperature decreases before reheating and 
the maximal temperature  $T_{\rm max}$ would typically be much larger than the reheating temperature $T_R$ \cite{Kolb:1990vq,Giudice:2000ex,Garcia:2020eof}.
One should keep in mind that thermalization is a complicated process such that the true maximal temperature is normally lower than that assumed in the instant thermalization approximation \cite{Mukaida:2015ria}. 
 This effect is, however, insignificant for our purposes.

 If the SM sector is produced by decay of a field other than the inflaton, the temperature scaling can be $a^0$ or  $a^{+|k|}$ meaning that 
 the maximal and reheating temperatures may be  close and even coincide,
 \begin{equation}
 T_{\rm max} \simeq T_R \;.
 \end{equation}
  A natural candidate for such a subdominant field is the right-handed neutrino $\nu_R$.
   It  can be produced via inflaton decay and subsequently decay into SM states. 
   The above calculation (\ref{rhoSM})  applied to the neutrino production $\phi \rightarrow \nu_R \nu_R$ implies
   \begin{eqnarray}
&& \rho_\nu \biggl\vert_{\phi^4}  = {3\over 2} \,\Gamma_\phi  H_0  M_{\rm Pl}^2\; {1\over a^2}\;, \\
&&  \rho_\nu  \biggl\vert_{\phi^2} = {6\over 5}\, \Gamma_\phi  H_0  M_{\rm Pl}^2\; {1\over a^{3/2}} \;,
\end{eqnarray}
   for the $\phi^4$ and $\phi^2$ local inflaton potentials, respectively.\footnote{In this work, $\phi^4$ and $\phi^2$ refer to the local behaviour of the inflaton potential around the minimum. The large field behaviour is expected to be very different from these since convex potentials are disfavored by the inflationary data.} $\rho_\nu$ is now the source for the SM radiation, so   
   using $\nu_R$ as $\chi$ in (\ref{system}), one finds
   \begin{equation}
   n-m=0
   \end{equation}
   as long as $\nu_R$ remains subdominant in the energy balance. 
    As a result, the SM bath temperature stays $constant$. In what follows, we study this scenario in more detail.

   We note that the SM temperature can also $increase$ before reheating. This happens if $n-m <0$, which can be realized in models where the SM sector is produced via decay of a sub-subdominant component
   in the energy density of the Universe. Indeed, the above considerations show that $n$ can be zero while $m>0$.
 
 \subsection{Example: SM sector production   via $\nu_R$ decay}
 
 Consider the possibility that the inflaton decays primarily into feebly interacting right-handed neutrinos, $\phi \rightarrow \nu_R \nu_R$. The relevant terms in the Lagrangian are
 \begin{equation}
 \Delta {\cal L}= y_\phi \, \phi \, \nu_R \nu_R + y_\nu\, H^c \bar \ell \nu_R +{\rm h.c.}\;,
 \end{equation}
 where the Yukawa coupling are  small, $y_\phi , y_\nu \ll 1$, and $H$ and $\ell$ are the Higgs and lepton doublet fields, respectively.
 Here we focus on the parameter regime where the direct coupling between the inflaton and the SM states is tiny such that $\phi \rightarrow hh$ is insignificant.
 This can be enforced 
 by an approximate $Z_4$ symmetry of the type    $\phi \rightarrow -\phi\; ; (\ell, \nu_R) \rightarrow i (\ell, \nu_R) $          \cite{DeRomeri:2020wng}. 
  
          \begin{figure}[h!] 
\includegraphics[scale=0.31]{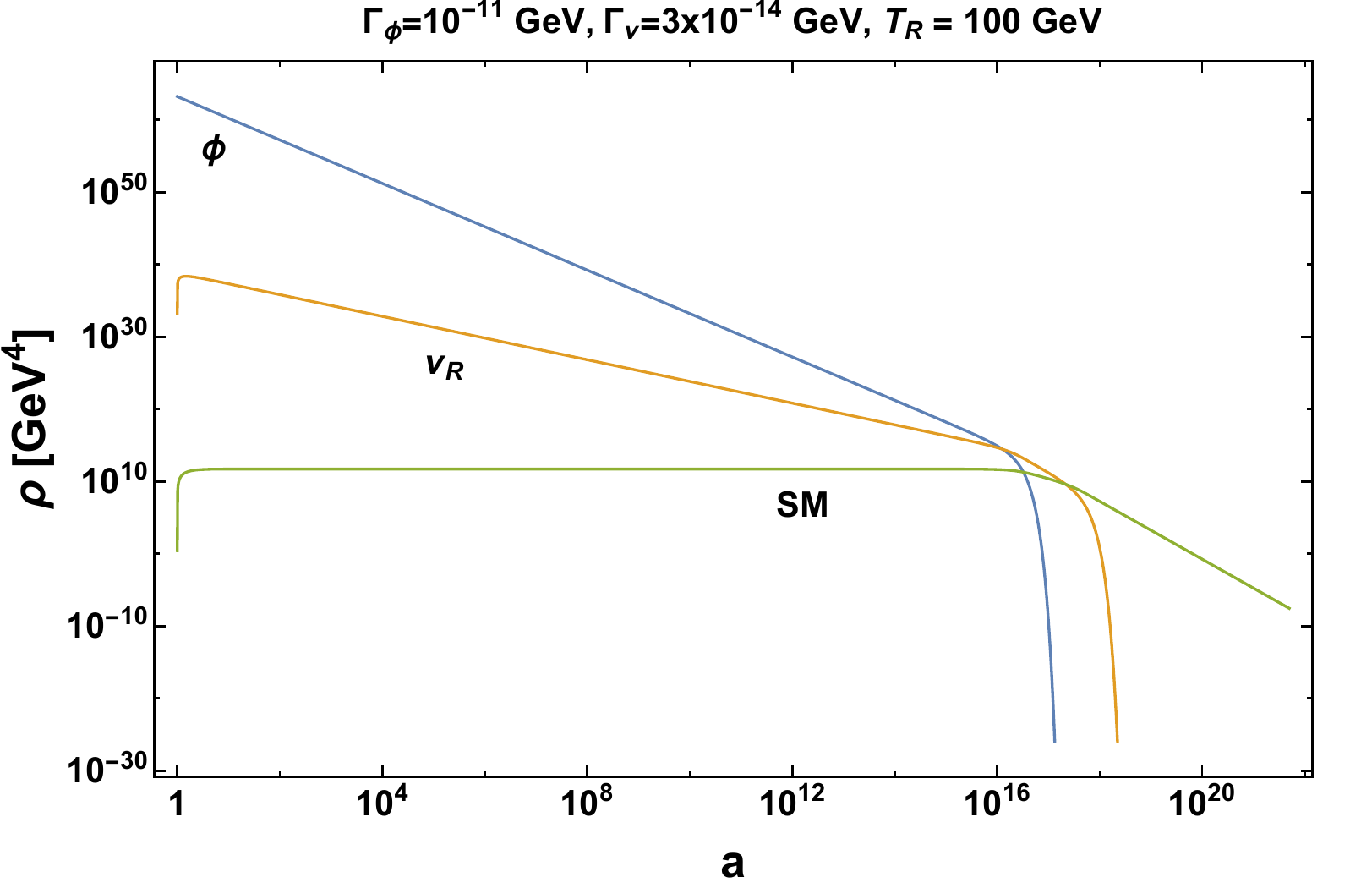}
\includegraphics[scale=0.31]{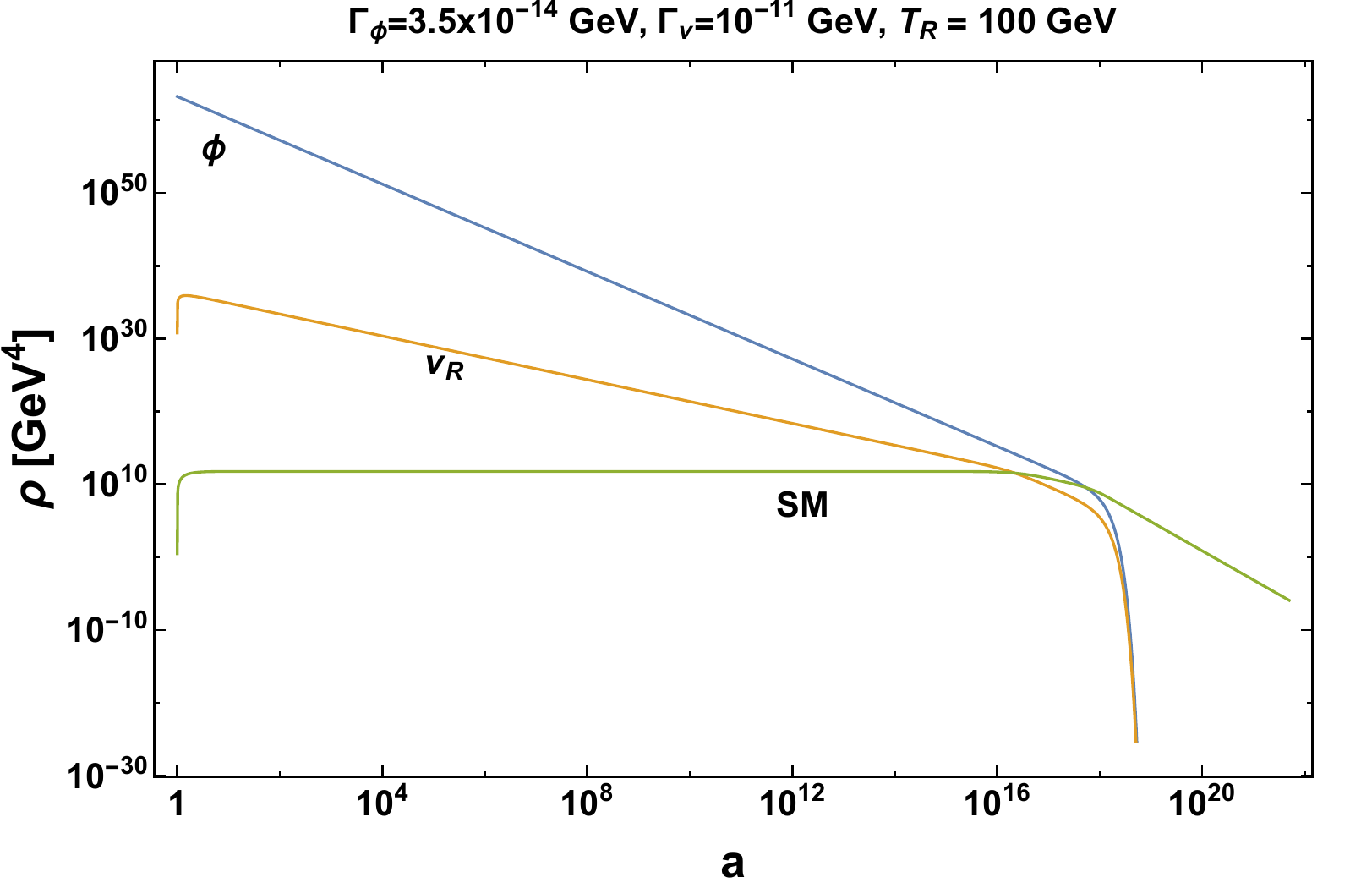}
 \\ \ \\
\includegraphics[scale=0.31]{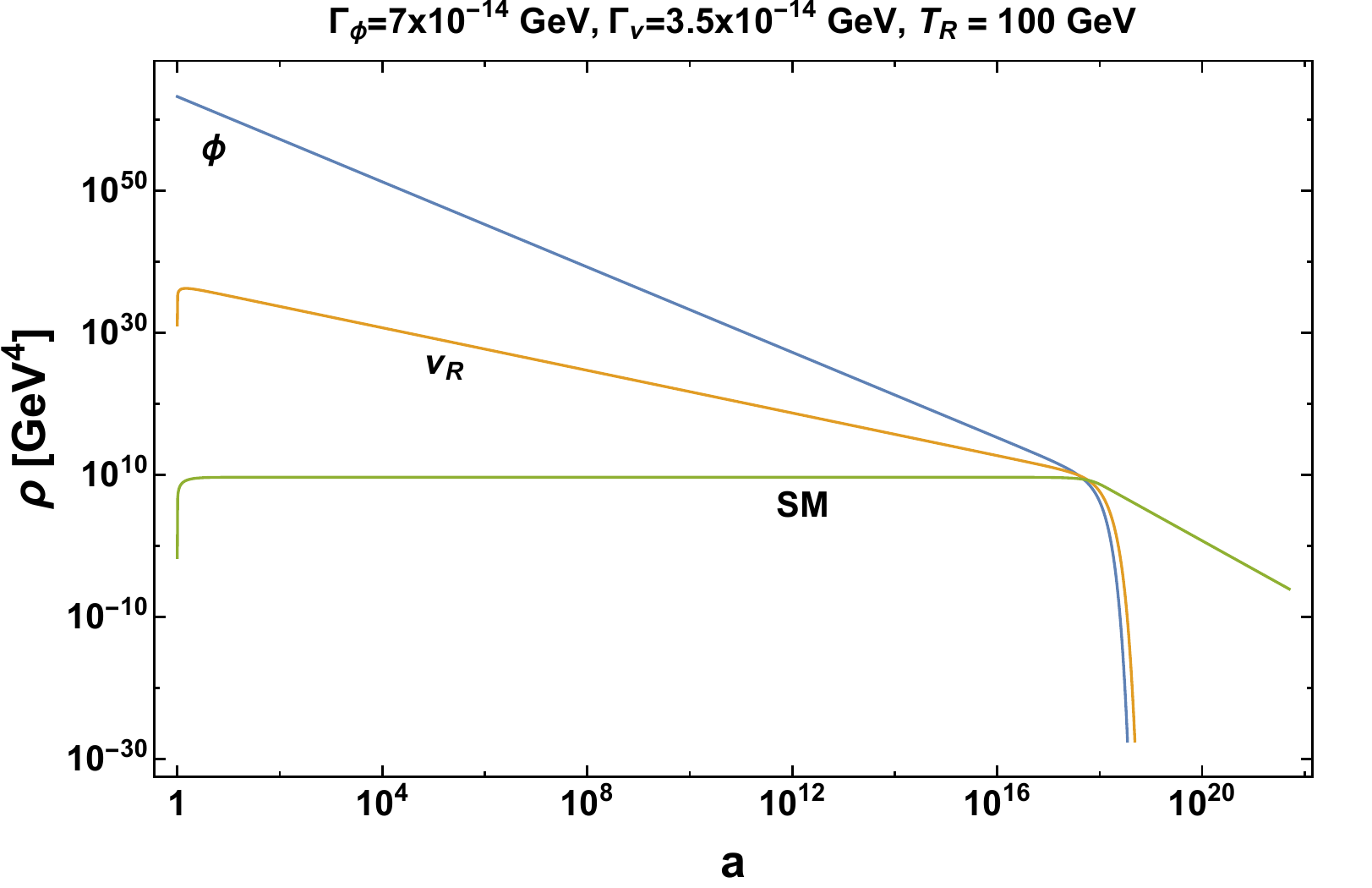}
\includegraphics[scale=0.34]{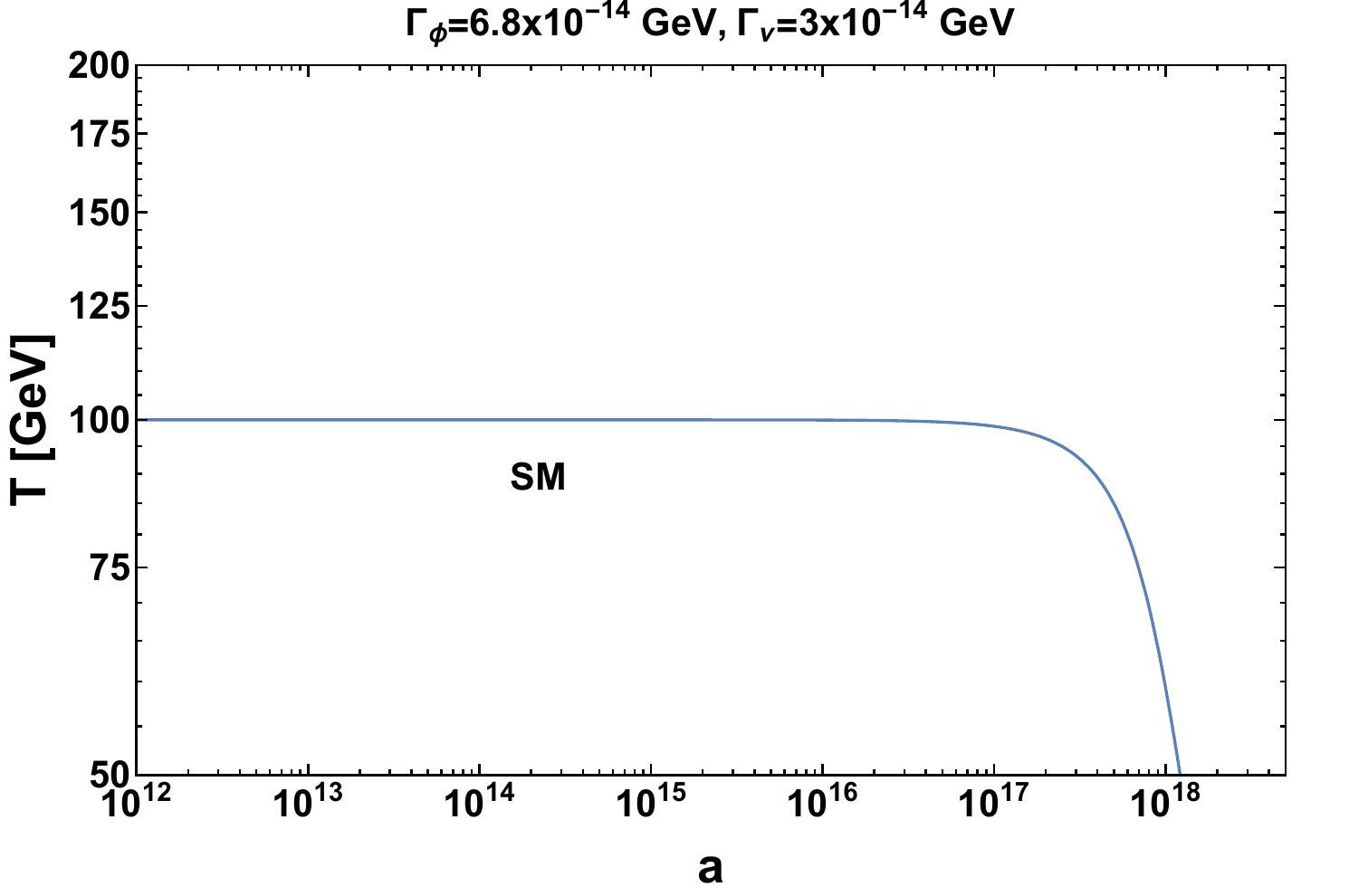}
\caption{ \label{1}
 Evolution of the energy density components and the SM sector temperature.
 {\it Top left:} $\Gamma_\phi \simeq  3.3\times10^2\, \Gamma_\nu$; {\it top right:} $\Gamma_\phi  \simeq 3.5\times 10^{-3}\, \Gamma_\nu$;
 {\it bottom left:} $\Gamma_\phi  \simeq 2 \,\Gamma_\nu$; {\it bottom right:} SM temperature evolution for $\Gamma_\phi  \simeq 2 \,\Gamma_\nu$.
}
\end{figure}

 The reheating process proceeds in two stages,
 \begin{equation}
 \phi \rightarrow \nu_R \nu_R ~~,~~ \nu_R \rightarrow {\rm SM} \;,
 \end{equation}
 where the final state in the $\nu_R$ decay can be $H \ell$ or lighter states, if the Higgs channel is forbidden kinematically. 
 We assume that $\nu_R$ is feebly coupled such that both decays take place at late times and the inverse processes can be neglected.
We also take the effective inflaton mass to be far above the $\nu_R$ mass, hence  
 the energy density of $\nu_R$ scales as radiation until late times. Given the feebleness of the right-handed neutrino interactions, it does not form a thermal bath.
 
 The energy density components satisfy the following equations
 \begin{eqnarray}
&& \dot \rho_\phi + 3H \rho_\phi = - \Gamma_\phi \rho_\phi ~,       \label{eq1} \\
&&  \dot \rho_\nu + 4H \rho_\nu =  \Gamma_\phi \rho_\phi - \Gamma_\nu \rho_\nu ~,       \label{eq2}\\
 &&  \dot \rho + 4H \rho =   \Gamma_\nu \rho_\nu ~, \label{eq3}    \\
 &&   \rho_\phi  + \rho_\nu + \rho = 3 H^2 M_{\rm Pl}^2 \;,     \label{eq4} 
\end{eqnarray}
where $\rho_\phi , \rho_\nu, \rho$ are the inflaton, $\nu_R$ and SM energy densities, respectively; $\Gamma_\phi $ and $\Gamma_\nu$ are the  decay rates of the corresponding energy densities.\footnote{The energy density decay widths $\Gamma_i$ are   such that possible $1+\omega_i$ factors, where $\omega_i$ is the equation of state coefficient, are absorbed in their definition. } The initial values for $\rho_\nu$ and $\rho$ are zero, and 
  $3H_0^2 M_{\rm Pl}^2= \rho_\phi(0) =V(\phi_0)$, where ``0'' refers to the end of inflation. Here we assume for definiteness  a non-relativistic inflaton, i.e.
$ V(\phi_0)= 1/2\, m_\phi^2 \phi^2_0$ with $\phi_0 \sim M_{\rm Pl}$, $H_0 \sim 10^{-5}M_{\rm Pl}$.
  
 As long as the decay widths are small and $ \Gamma_\phi \rho_\phi \gg  \Gamma_\nu \rho_\nu$, 
 the energy densities are hierarchical and 
 we recover the scaling behaviour of the solution described earlier:
 \begin{equation}
 \rho_\phi \propto a^{-3} ~,~  \rho_\nu \propto a^{-3/2} ~,~ \rho \propto a^0 \;
 \end{equation}
 at $a>{\cal O}(1)$.
   Furthermore, even if $\Gamma_\nu $  
 is time-dependent, the SM energy density can be computed via
 \begin{equation}
\rho(a) =   {1\over a^4}    \int_1^a da \;a^3 \, \Gamma_\nu \,{\rho_\nu \over H} \,.
\end{equation}
 This expression makes it explicit that if $\rho_\nu$ scales as $H$ and $\Gamma_\nu=\;$const,  the SM energy density remains constant. The latter  can also $increase$ in time  if  
 $\Gamma_\nu \,{\rho_\nu / H} $ grows, which could be due to the increase of ${\rho_\nu / H}$ or broadening of the neutrino width.
 
\begin{figure}[h!] 
        \begin{center}
\includegraphics[scale=0.35]{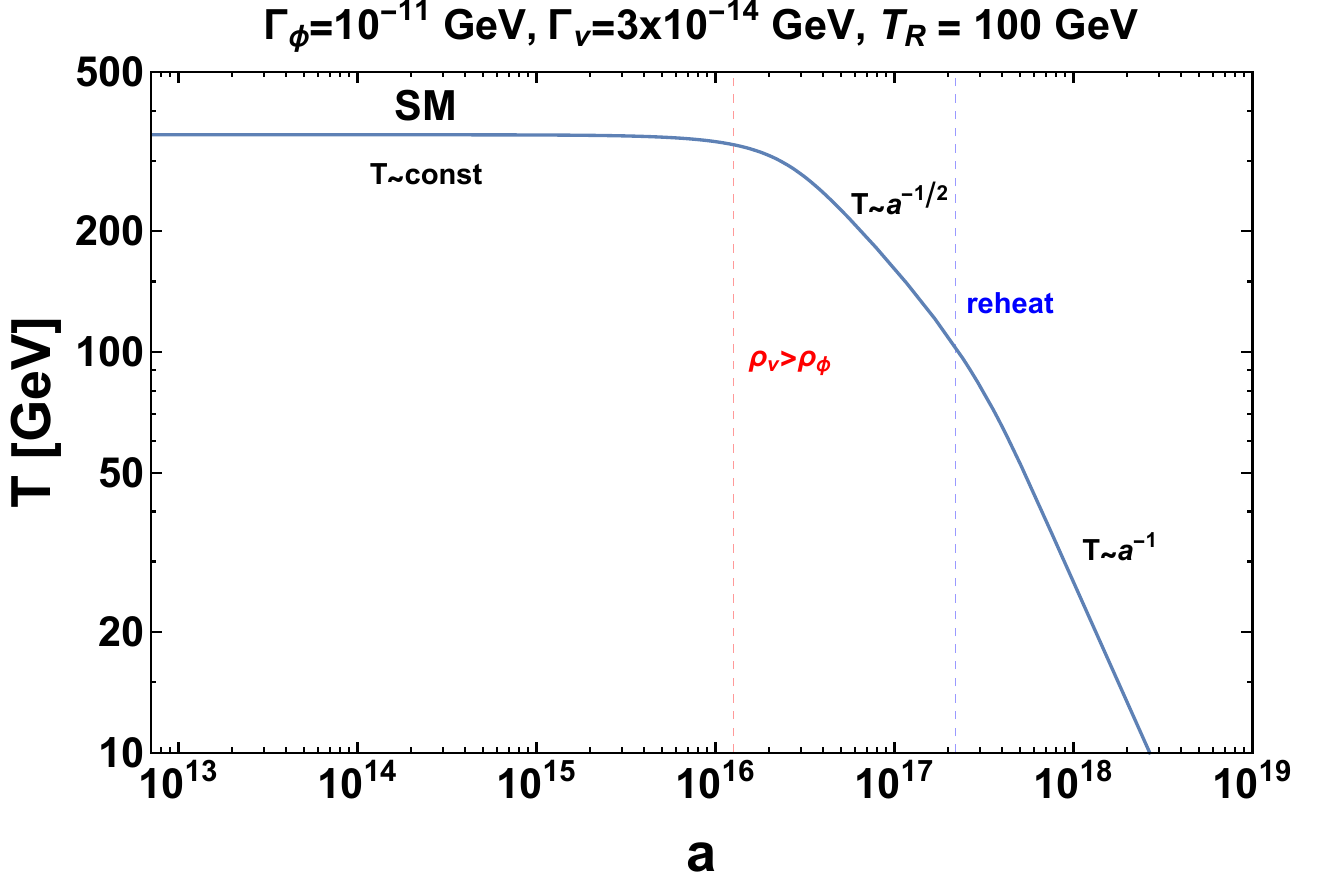}
\includegraphics[scale=0.35]{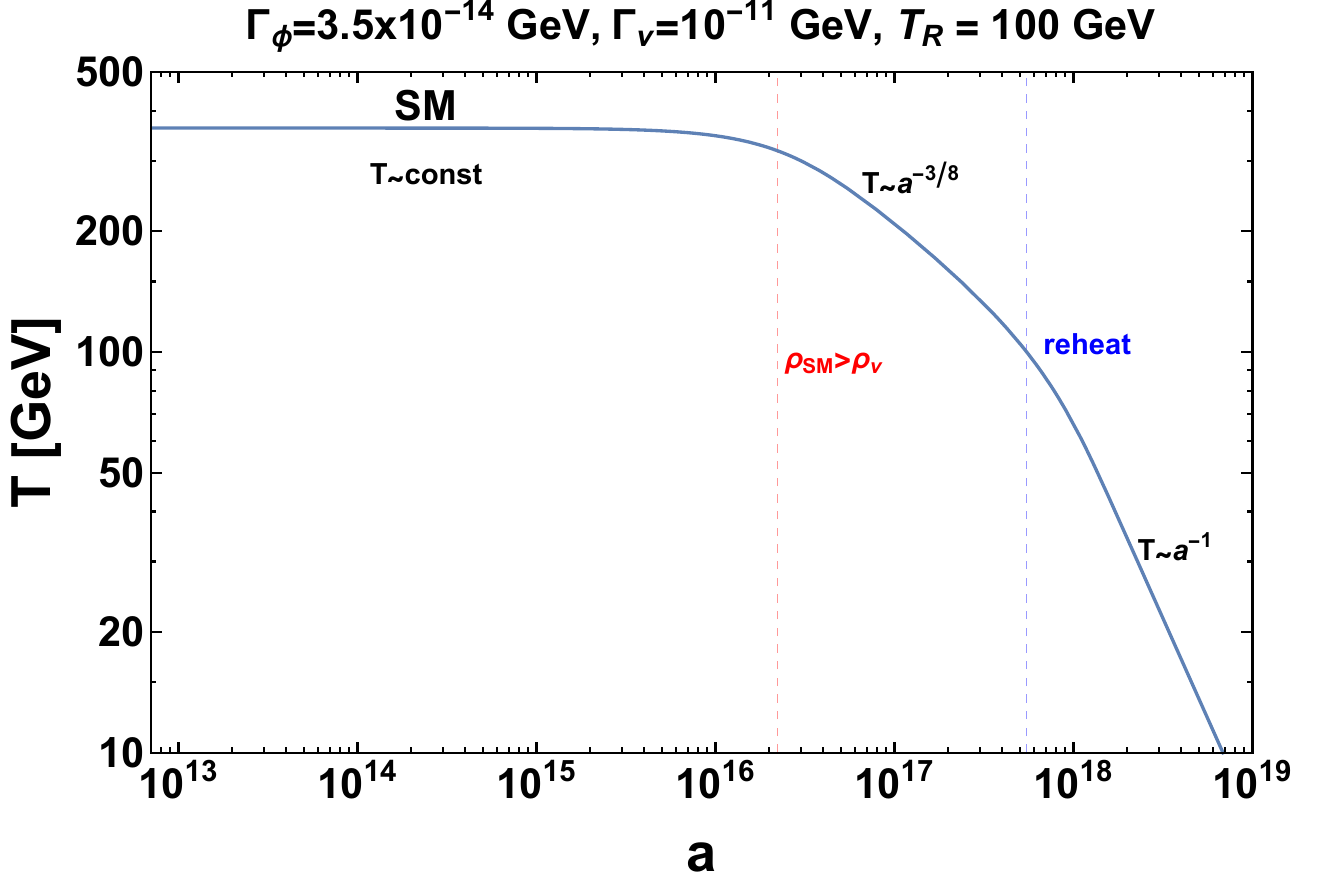}
\caption{ \label{T-plots}
 Temperature evolution for $T_{\rm max} > T_R$.  The parameters are as in Fig.\,\ref{1}  (upper panels) with the focus on the vicinity 
 of the reheating region.}
\end{center}
\end{figure}

 At late times, the different energy densities start approaching each other and their hierarchical structure gets lost. In this case, the full numerical solution is necessary (Fig.\,\ref{1}). Below, we distinguish the cases 
 of 
 $\Gamma_\phi  $   and $ \Gamma_\nu $ being vastly different,      and being of similar size.
 \\ \ \\
 {\underline{$\Gamma_\phi  \gg \Gamma_\nu$\,.}}
 The inflaton field decays into the right-handed neutrinos and drops out of the energy balance 
 when $H \sim\Gamma_\phi$. Until SM reheating, the dominant role is taken by relativistic $\nu_R$. In this period, the SM temperature decreases as $T \propto a^{-1/2}$ (Fig.\,\ref{T-plots}). 
 Reheating occurs when $H \sim\Gamma_\nu$. 
 \\ \ \\
  {\underline{$\Gamma_\phi  \ll \Gamma_\nu$\,.}}
  As long as $ \Gamma_\phi \rho_\phi \gg \Gamma_\nu \rho_\nu$, the energy components evolve as in the previous case.
  The transition occurs when $ \Gamma_\phi \rho_\phi \sim \Gamma_\nu \rho_\nu$. The neutrino energy density $ \rho_\nu$ starts decreasing faster ($\rho_\nu \propto \rho_\phi \propto a^{-3}$), although it 
  does not decay exponentially quickly due to continuous replenishment from the inflaton sector. 
  The SM sector takes over the subleading role in the energy balance until
  $H \sim\Gamma_\phi$, when both $\nu_R$ and $\phi$ decay exponentially quickly.
  The SM temperature decreases in this period as $T \propto a^{-3/8}$ (Fig.\,\ref{T-plots}), so again 
  the maximal temperature exceeds the reheating temperature.
  \\ \ \\
  {\underline{$\Gamma_\phi  \sim \Gamma_\nu$\,.}}
  This parameter choice realizes the possibility $T_{\rm max} \simeq T_R$. Since the inflaton and  neutrinos decay at the same time, the SM sector takes over the energy balance immediately thereafter.
 As Fig.\,\ref{1} (bottom right) demonstrates, the temperature evolution can be approximated by a constant followed by a power law decrease $T \propto a^{-1}$.
   \\ \ \\
   We therefore conclude that, in the simple framework where the SM sector is produced by the right-handed neutrino decay,  
the maximal and reheating temperatures do not differ significantly if   the inflaton and $\nu_R$ decay widths are similar.
Their approximate equality  $T_{\rm max} \simeq T_R$ is realized for $\Gamma_\phi  \simeq \Gamma_\nu$.

  One should also note that certain thermal effects can play a role in flattening the $T(a)$ dependence prior to reheating. In particular, if $T$ is far above the neutrino mass $M_\nu$, which we keep as a free parameter,   the neutrino decay 
  into the SM final states is blocked kinematically. This can be accounted for by replacing $  \Gamma_\nu      \rightarrow   \Gamma_\nu   \, \theta(\rho_\star - \rho)$ in the evolution equations.
  Here $\rho_\star$ is proportional to $M_\nu^4$ and for values of $\rho_\star \sim T_R^4$, the SM temperature stays constant till reheating even if $\Gamma_\phi \gg \Gamma_\nu$.

  Finally, let us estimate the size of the couplings required for a low reheating temperature. 
  The relation $\Gamma_\phi  \sim \Gamma_\nu$ implies 
  \begin{equation}
  {y_\nu \over y_\phi} \sim \sqrt{m_\phi \over M_\nu} \;,
  \end{equation}
  where $m_\phi$ and $M_\nu$ are the inflaton and right-handed neutrino masses, respectively. This is because both $\phi \rightarrow \nu_R \nu_R$ and $\nu_R \rightarrow H \ell$ decay widths scale as the initial particle mass times the coupling squared. The typical parameter values in Fig.\,\ref{1} are $T_R \sim 100$ GeV and $\Gamma_i \sim 10^{-14}$ GeV, which for $m_\phi \sim 10^{13}$ GeV implies
  $y_\phi \sim 10^{-13}$. Assuming $M_\nu \sim 1$ TeV, the neutrino Yukawa coupling is then $y_\nu \sim 10^{-8}$, consistent with its non-thermalization \cite{DeRomeri:2020wng}.  (For light $\nu_R$, its decay can be loop-suppressed.)
  The inflaton coupling appears particularly small, although it is not surprising in inflationary cosmology given the flatness of the inflaton potential. The smallness of the couplings in the neutrino sector is also 
  not unexpected, hence one may find the above scenario plausible, although the parameter choice can only be motivated in a more fundamental theory.

  These estimates also show that the direct decay $\phi \rightarrow\;$SM would be highly suppressed in this model. The inflaton coupling to the SM fields is generated at one loop, 
  e.g. the coupling to the Higgs bilinear $H^\dagger H$ is proportional to the
  loop factor$\,\times\, y_\phi y_\nu^2 \, M_\nu$, where the last factor is required by the $Z_4$ symmetry  \cite{DeRomeri:2020wng}   breaking. This  makes the decay rate of  $\phi \rightarrow\;$SM suppressed by about  56 orders of magnitude compared to that of 
     $\phi \rightarrow\nu_R \nu_R$. More generally, the SM energy density  generated via the cascade $\phi \rightarrow \nu_R \rightarrow\;$SM scales as $\Gamma_\phi \Gamma_\nu$, whereas that 
     from the direct decay $\phi \rightarrow \;$SM scales as  $\Gamma_\phi \Gamma_\nu^2$. Therefore, the latter can be made highly suppressed by decreasing  $\Gamma_\nu$.

  \section{Dark matter freeze-in production at stronger coupling}
  
  Let us consider freeze-in dark matter production in the class of models described in the previous section.
  The SM sector temperature has never been very high in this framework.  The maximal and reheating temperatures are of similar size and possibly coincide.
  Before reheating, the Universe undergoes a long period of inflaton matter-dominated expansion  which dilutes the relativistic relics generated right after inflation.
  Hence, one may focus entirely on thermal production of dark matter. 
  
   We
  are interested in freeze-in at larger couplings \cite{Cosme:2023xpa}, so we 
   take the DM mass to be much larger than the SM temperature at any stage of the Universe evolution. On the other hand, the temperature profile is model-dependent
   and can be highly non-trivial in a multi-component Universe. We thus focus on a simplified case motivated in the previous section. Specifically, we 
   assume that the relevant stage in the Universe evolution can be approximated by the period of constant SM temperature, followed by a power-law decrease $T\propto a^{-l}$ as
   displayed in Fig.\,\ref{1} (bottom right).
   We focus on the parameter regime in which the DM production via direct decay of the inflaton or the right-handed neutrino is highly suppressed or forbidden.\footnote{
   In the example of the previous section, the inflaton decay into DM is highly suppressed by powers of small couplings as well as $M_\nu/m_\phi,\,m_h/m_\phi$, while the $\nu_R$ decay into DM plus leptons 
would  be forbidden kinematically if $M_\nu < 2 m_{\rm DM} $.}

   Consider, for definiteness, real scalar dark matter $s$ of mass $m_s $ which couples to the SM sector through  the Higgs portal \cite{Silveira:1985rk,Patt:2006fw,Lebedev:2021xey},
  \begin{equation}
 V(s) = {1\over 2}  \lambda_{hs }s^2 H^\dagger H + {1\over 2} m_s^2 s^2 \;,
 \end{equation}
 where $ \lambda_{hs }$ is the Higgs portal coupling.
For $T \ll m_s$, dark matter is produced 
   via scattering of energetic  SM quanta from the ``Boltzmann tail''.
   The leading     process is the Higgs  and vector boson pair annihilation into pairs of $s$. For $m_s \gg m_h$,  all these modes can be accounted for by 
 using four effective  Higgs degrees of freedom, according to 
  the Goldstone  equivalence theorem.

  The dark matter number density $n$ satisfies the Boltzmann equation
  \begin{equation}
\dot n + 3Hn = \Gamma_{h_i h_i \rightarrow ss}  -   \Gamma_{ss\rightarrow h_i h_i}  \;,
\end{equation}
where $h_i$ represents 4 Higgs d.o.f. at high energies of order $m_s \gg m_h$; 
$\Gamma_{h_i h_i \rightarrow ss}  $ and  $ \Gamma_{ss\rightarrow h_i h_i}$  are the DM production and annihilation rates per unit volume, respectively. 

Consider first the pure freeze-in regime, in which the annihilation term can be neglected.  Since the initial state particles have energies far above the temperature, their distribution is given by the ``Boltzmann tail''.
Then, the production rate per unit volume is \cite{Cosme:2023xpa}
\begin{equation}
  \Gamma_{h_i h_i \rightarrow ss}   \simeq  {\lambda_{hs}^2 \, T^3  m_s \over 2^7 \pi^4} \,   e^{-2m_s/T} \;,
    \end{equation}
 for 4 Higgs d.o.f.  
Integrating the Boltzmann equation, we get
 \begin{equation}
 a^3 n =  H_0^{-1} \; \int da \; a^{m+2}\,  \Gamma_{h_i h_i \rightarrow ss}(a)\;. 
 \end{equation}
Here we use the scaling $H= H_0 / a^m$, as before. Note that one cannot trade the integration variable $a$ for $T$ since $T(a)$ can be constant. Using the explicit form of the reaction rate, we find
\begin{equation}
 a^3 n =  {\lambda_{hs}^2  \, m_s \over 2^7 \pi^4 H_0} 
   \; \int da \; a^{m+2}\, T^3(a) \,e^{-2m_s/T(a)} \;. 
 \end{equation}
Consider now two representative  $T(a)$ functions. 

\subsection{Constant $T(a)=T_{\rm max}$}

This scaling applies when the SM sector is a sub-subdominant component in the energy density of the Universe.
The integral is trivial and the total particle number is
\begin{equation}
 a^3 n =  {\lambda_{hs}^2  \, m_s \over 2^7 \pi^4 H_0} \, {a^{m+3}\over m+3 }\; T^3_{\rm max} \,e^{-2m_s/T_{\rm max}}
 \label{Tconst}
  \end{equation}
for $a\gg a_0=1$.   
 We see that the particle number  (and the density) grows with $a$ and is dominated by the late time contributions at largest $a$.\footnote{A similar result applies to the case when $T$ increases in time: the particle number is dominated by the largest $a$ contribution.}

  \subsection{$T(a)\propto a^{-l} $}

During and after reheating, the SM temperature decreases. We may approximate it    by a power law $T(a)\propto a^{-l} $, with $l$ time-dependent but  approximately constant during each  period of interest.
 Let us take 
\begin{equation}
T(a) = T_{\rm max} \; \left( {a_{\rm max} \over a}\right)^l  \;,
\end{equation}
where $a_{\rm max}$  corresponds to the point where the temperature starts decreasing,
and compute the particle number generated from $a_{\rm max}$ on.
Using the approximation
\begin{equation}
\int_{z_0}^\infty  dz \, z^k e^{-z} \simeq z_0^k \,e^{-z_0} 
\end{equation}
for $z_0 \gg 1$, obtained using integration by parts, we get
\begin{equation}
 a^3 n \simeq   {\lambda_{hs}^2 \, m_s \over 2^7 \pi^4 H_0} \, {a^{m+3}_{\rm max}\over l }\;  {T_{\rm max}\over 2 m_s}  \;T^3_{\rm max} \,e^{-2m_s/T_{\rm max}}\;.
 \label{T-l}
  \end{equation}
  This expression assumes $2m_s/T_{\rm max} \gg 1$. In contrast to the previous case, the particle number is dominated by the early time contribution. This is due to the exponential suppression of
  particle production at lower temperatures.
  
  Both (\ref{Tconst}) and (\ref{T-l}) can be rewritten in terms of $H(a)$. The time dependence of the particle density is captured by
  \begin{equation}
  n(a)\biggl\vert_{T={\rm const}} \propto  {1\over H(a)} ~~,~~
  n(a)\biggl\vert_{T \propto a^{-l}} \propto  {1\over a^3 \; H(a_{\rm max})  }  \;.
  \end{equation}

  \subsection{Total particle number}
  
  Let us now combine the two contributions.
  Defining $a_{\rm max}$ as 
 the transition point from one scaling law to the other, we obtain the  
       total DM particle number produced by the SM thermal bath,
  \begin{equation}
a^3 n \biggl\vert_{\rm tot}  =  a^3_{\rm max} n(a_{\rm max}) \biggl\vert_{T={\rm const}} \times  \left(  1+ {m+3 \over l } \;   {T_{\rm max}\over 2 m_s}  \right) \;,
\label{a3n-tot}
   \end{equation}
   where $ a^3_{\rm max} n(a_{\rm max}) \bigl\vert_{T={\rm const}}$ is found from (\ref{Tconst}).
   DM production is most efficient in the vicinity of $a=a_{\rm max} $ with the flat part of the temperature evolution giving the main contribution since 
   ${T_{\rm max}\over 2 m_s} \ll 1$. We note that the $l$-dependence of the result is very mild: $l$ only affects the correction.
   
The consequent constraint on the model  is  formulated in terms of the relic abundance of the $s$-particles,
   \begin{equation}
Y = {n \over s_{\rm SM}} ~~,~~  s_{\rm SM} = {2\pi^2 g_*\over 45} \, T^3   \;,
\end{equation}
where $g_*$ is the number of the SM d.o.f.  and $n$ is the number density of the $s$-quanta. The observational constraint on the DM abundance is 
$Y_{\rm obs}= 4.4 \times 10^{-10} \; \left(   {{\rm GeV}\over m_s}  \right)$ \cite{Planck:2015fie}. 
   
   Assuming that $T_{\rm max}=T_R$, the Hubble rate at $a=a_{\rm max} $ is determined by the SM temperature, $H_{\rm max} \propto T_{\rm max}^2$ and $l=1$.
   We thus get
    \begin{equation}
Y \simeq {\sqrt{90} \, 45 \over  2^{8} \pi^7 \, g_*^{3/2}  }             \,
{\lambda_{hs}^2 \, m_s \,M_{\rm Pl}   \;  e^{-2m_s/T_{\rm max}}   \over  T_{\rm max}^2}  \,
\left(     {1\over m+3} +   {T_{\rm max}\over 2 m_s}   \right) \;.
\label{Ytot}
\end{equation}
 To obtain this expression, we have estimated $Y$ at $T= T_{\rm max}$, after which the abundance remains approximately constant in the pure freeze-in regime. 
 The correct DM relic abundance is reproduced for $\lambda_{hs} \sim 10^{-11} \,e^{m_s/ T_{\rm max}} \, \sqrt{m_s/ {\rm GeV}}$. The coupling can be as large as order one if
 $m_s \gg T_{\rm max}$. On the other hand, models with a high reheating temperature require $\lambda_{hs} \sim 10^{-11}$ \cite{Yaguna:2011qn,Lebedev:2019ton}.

\begin{figure}[h!] 
        \begin{center}
\includegraphics[scale=0.90]{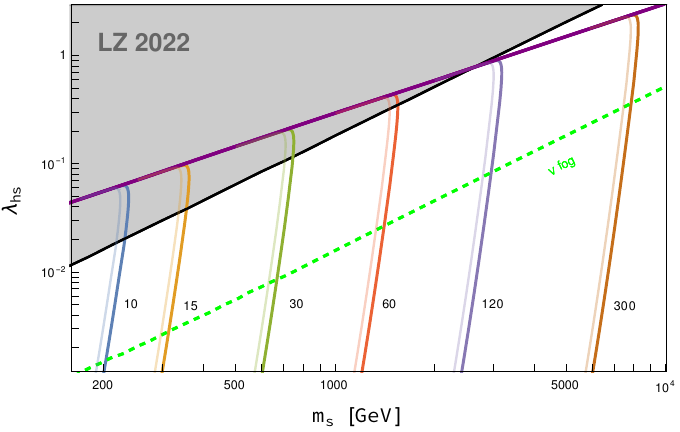}
\caption{ \label{par-space}
 Parameter space of the Higgs portal DM model. Along the colored lines, the correct relic density is reproduced for a given $T_R = T_{\rm max}$  (in GeV). The thin lines correspond to the calculation presented in \cite{Cosme:2023xpa}, while the thick lines are obtained in the UV complete model of the current work.  The purple line corresponds to the thermal DM abundance. The shaded area is excluded by direct DM detection (LZ 2022)
 \cite{LZ:2022ufs} and the dashed line shows the ``neutrino fog''. }
\end{center}
\end{figure}

   In our previous work  \cite{Cosme:2023xpa}, we resorted  to  a low energy effective approach without specifying the full cosmological history. In particular, we assumed zero particle abundance before reheating, 
   which amounts to computing particle production from the ${T \propto a^{-l}}$ region only.
   Taking $m=3/2$ corresponding to inflaton matter domination prior to reheating, we find that ignoring ``prehistory'' leads to underestimating the particle number   by about a factor of 10. 
   To reproduce the correct DM relic abundance, the effective approach required   
   $2m_s/T_R \sim 40$ for typical parameter values. Since $Y$ scales approximately as  $e^{-2m_s/T_R}$,
   the 10-fold increase in $Y$ can be compensated by a 5\% shift in $T_R$.
   Hence, the full computation leads to a  $T_R$ correction of order 5\%,
    \begin{equation}
T_R \rightarrow 0.95 \times T_R \;.
   \end{equation}
   Equivalently, for a fixed $T_R$, the DM mass $m_s$ increases  by about 5\%. Hence, 
  the parameter space plot remains largely unchanged. This is shown in Fig.\,\ref{par-space}.
   The thin and thick colored lines correspond to the correct relic density computed within the effective approach of our previous work \cite{Cosme:2023xpa} and using the UV complete model presented here, respectively.

   At larger couplings, the annihilation effect becomes significant. The corresponding reaction rate per unit volume is given by \cite{Cosme:2023xpa}
    \begin{equation} 
  \Gamma (ss \rightarrow h_ih_i)  =  \sigma (ss \rightarrow h_ih_i ) v_r \;n^2~~, ~~ \sigma (ss \rightarrow h_ih_i ) v_r=  4 \times {\lambda_{hs}^2 \over {64 \pi m_s^2} }
  \end{equation}
for $m_s^2 \gg m_h^2$ and 4 Higgs d.o.f. Here $v_r$ is the relative particle velocity. 
As one increases the coupling further, the system thermalizes         \cite{Cosme:2023xpa,Silva-Malpartida:2023yks}
and the  resulting DM relic abundance is given by the purple line in Fig.\,\ref{par-space}.

\begin{figure}[h!] 
        \begin{center}
\includegraphics[scale=0.870]{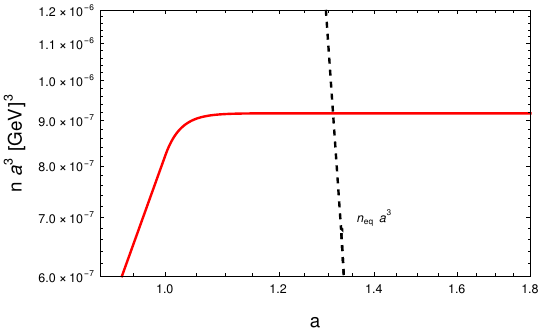}
\includegraphics[scale=0.83]{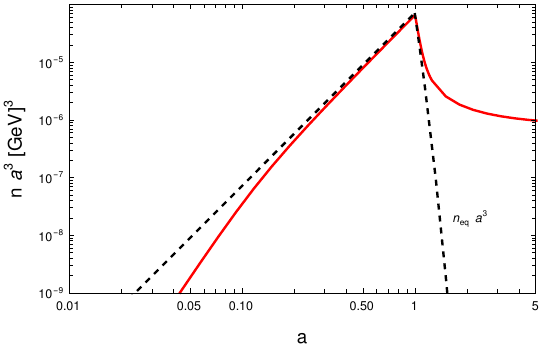}
\caption{ \label{n}
 Dark matter particle number evolution. 
 $n_{\rm eq}$ represents  the thermal particle density at the SM sector temperature $T$, which stays constant before $a=1$ and decreases as $1/a$ after that.
 {\it Left:} pure freeze-in regime ($\lambda_{hs}=10^{-3}$, $m_s=602.5\;$GeV, $T_{\rm max}=30\;$GeV).
 {\it Right:} DM thermalization and freeze-out ($\lambda_{hs}=0.195$, $m_s=651\;$GeV, $T_{\rm max}=30\;$GeV).}
\end{center}
\end{figure}

     Fig.\,\ref{n} illustrates  the process of thermalization with our $T(a)$ function. At low couplings, the DM particle number differs from its equilibrium value at all stages, apart from one point. 
     For larger couplings,
     the DM density  reaches the equilibrium value somewhat  before $a=a_{\rm max}$, which is followed by DM freeze-out. For convenience, we redefine the scale factor in the plot as $a/a_{\rm max} \rightarrow a$
     such that the SM temperature starts decreasing at $a=1$ and $T_R=T(1)$.

 \subsection{ $T_{\rm max} > T_R $ case}
  
  More generally,  $T_{\rm max}$ and $T_R $ can differ. For instance, if $\Gamma_\phi \gg \Gamma_\nu $  or $\Gamma_\phi \ll \Gamma_\nu $, the maximal temperature exceeds the reheating temperature. 
  In this case, the flat $T=\;$const stage, during which the inflaton field dominates, is followed by a period of a mild temperature decrease $T\propto a^{-l}$ with $l<1$. During this stage, the right-handed neutrinos
  take over the energy balance. As these decay, the SM bath comes to dominate and $l$ becomes 1 signifying reheating. Thus, $T_R $ can be significantly lower than $T_{\rm max}$. 
  
  Consider the pure freeze-in regime. It is clear from the above calculations that  dark matter is mostly produced at the point $a=a_{\rm max}$.
  Hence, Eq.\,\ref{a3n-tot} applies to this case as well. However, the resulting DM abundance changes   since the SM entropy gets produced until reheating, after which $Y $ remains approximately constant.
   The result can be conveniently expressed as a correction factor to (\ref{Ytot}).  Let us parametrize the scaling of the Hubble rate between $a_{\rm max}$ and $a_R$ as $H \propto a^{-m'}$, such that
  $H$ evolves with $a$ according to 
  \begin{equation}
1  \;\stackrel{a^{-m}}  \longrightarrow     \;    a_{\rm max}   \;  \stackrel{ a^{-m'}}  \longrightarrow  \; a_R      \;\stackrel{a^{-2}}  \longrightarrow    \infty   \;.
\end{equation} 
   Since the DM particle number does not change between $a_{\rm max}$ and $a_R$, we obtain the following correction factor to the result (\ref{Ytot}):
  \begin{equation}
  Y\simeq  Y\biggl\vert_{T_{\rm max}=T_R} \times \left(    {T_R \over T_{\rm max}}  \right)^{{m'+3 \over l}-5} \;.
  \label{rescaling}
  \end{equation}
  For example, in the case of radiation domination, $m'=2$ and $l=1/2$. Hence, the correction factor is $(T_R/T_{\rm max})^5$. 
 Although this factor can be substantial, it does not change the parameter space picture drastically: taking $T_{\rm max} =10 \,T_R$ results in about a 25\% shift in $m_s$. 
  
  Freeze-in production at stronger coupling implies that the $maximal$ temperature is far below the DM mass. In this case, particle  production  
  is always Boltzmann-suppressed and of freeze-in type. This condition is typically violated if the inflaton decays directly into the SM states, implying a high $T_{\rm max}$.
  The dynamics of particle production would be more complicated in this case: depending on further details, DM may thermalize and  freeze out before Boltzmann-suppressed freeze-in 
  becomes operative.

Many of our results also apply  if the SM sector is produced by a         sub-subdominant component in the energy density of the Universe.\footnote{This refers to the energy balance immediately after inflation,
i.e. $n-m<0$ in (\ref{rhoSM}). Shortly before reheating, this component may dominate the energy density.}
 In this case, the SM sector temperature increases according to a power law
until the energy balance changes and the  evolution of $T$ flattens, followed by its decrease. DM production is most efficient at the end of the flat $T$ period, hence the above considerations remain largely
valid.

In summary, the main results of our work are presented in Fig.\,\ref{par-space}. The direct detection experiments already probe freeze-in DM up to 3 TeV. As their sensitivity increases, they will cover 
further consistent freeze-in models with higher reheating temperature or lower couplings. The difference between the full model results with $T_R = T_{\rm max}$ and those using the instant reheating approximation is small in terms of the $\{\lambda_{hs}, m_s\}$ parameter space.  If $T_{\rm max}$ is significantly above $T_R$, the $\{\lambda_{hs}, m_s\}$ relation receives a correction 
due to  (\ref{rescaling}): 
\begin{equation}
\lambda_{hs} \rightarrow \lambda_{hs} \times \left(    {T_{\rm max} \over T_{R}}  \right)^{{m'+3 -5l\over 2l}}\;,
\label{lambda-rescale}
\end{equation}
as long as $m_s \gg T_{\rm max}$.
Due to the exponential 
mass-dependence, this results in a modest shift of the $\{\lambda_{hs}, m_s\}$ curve at fixed $T_R$ to smaller  masses. 
For the same reason, model-dependent pre-thermalization DM production via scattering of the high energy quanta against the SM thermal bath \cite{Harigaya:2014waa,Mukaida:2022bbo} is not expected to affect our results significantly.

At low $m_s$,  the direct   detection sensitivity is normally superseded by that of the LHC via the   constraint on the  invisible Higgs decay. This also applies to the Higgs portal model with low $T_R$ \cite{Bringmann:2021sth}. Indirect detection constraints are weak in the Higgs portal models \cite{Lebedev:2021xey}. 
Let us note that there exist other examples of freeze-in models, although with a  more complicated dark sector, that can potentially be probed by a combination of experiments
\cite{Hambye:2018dpi,An:2020tcg,Bhattiprolu:2022sdd}. For example, milli-charged DM with a light mediator is already constrained by direct DM detection \cite{Hambye:2018dpi}.
Some freeze-in models assuming early matter-domination can also be probed via displaced vertices at colliders \cite{Calibbi:2021fld}.

   Finally, we note that baryogenesis in our framework can be realized via low-temperature Affleck-Dine mechanism \cite{Affleck:1984fy}. It requires the existence of a scalar condensate, which can naturally be generated, for instance,
in  the pseudo-Goldstone direction in the scalar potential   \cite{Harigaya:2019emn}. Further constraints on specific realizations of our scenario can be obtained by assuming a particular inflationary potential,
as detailed in 
 \cite{Becker:2023tvd}.

\section{Conclusion}

Dark matter freeze-in       at stronger coupling is a well motivated scenario, which addresses the problem of initial conditions in conventional freeze-in models.
It assumes
a low reheating temperature allowing for dilution of gravitationally produced relics and making DM production Boltzmann-suppressed. 
As a result, the DM coupling to the SM fields is allowed to be ${\cal O}(1)$ rendering it potentially observable in direct detection experiments \cite{Cosme:2023xpa}.
On the other hand, the  DM relic abundance  in this framework is sensitive  to
the thermal history of the Standard Model sector   and, in particular, to the relation between the maximal and reheating temperatures.

 In this work, we have studied a class of models, in which       the maximal and reheating temperatures are in the same ballpark, and can even coincide.
 This is the case when  the SM sector is produced via decay of a subdominant component in the energy density of the Universe. The role of this component can be played by feebly interacting right-handed neutrinos $\nu_R$. If these couple to the inflaton much stronger than the SM fields do,  the inflaton decays  predominantly into pairs of $\nu_R$, which subsequently produce the SM sector. 
 In this case, the SM bath temperature stays constant over       cosmological times prior to reheating, explaining the proximity of     the maximal and reheating temperatures.                         

Since the temperature evolution is known in this framework, the dark matter abundance can be calculated reliably. We find that the resulting constraint on the coupling vs mass is  very
 close to that  computed in the instant reheating approximation \cite{Cosme:2023xpa} as long as  $T_R \simeq T_{\rm max}$. If the maximal and reheating temperatures differ, the coupling receives a rescaling factor 
 (\ref{lambda-rescale}).

An attractive feature of freeze-in models with stronger coupling is that they are currently being probed by direct DM detection experiments. As the search sensitivity  improves \cite{XENON:2020kmp,DARWIN:2016hyl}, further models with higher $T_R$
or lower couplings will be explored. In addition, the LHC can place a complementary constraint via precise measurements of the Higgs invisible decay.
 All in all, dark matter freeze-in       at stronger coupling provides us with an interesting alternative to traditional freeze-in or freeze-out models.
 \\ \ \\
 {\bf Acknowledgements.}
  C.C. is supported by the FCT - Funda\c{c}\~{a}o para a Ci\^{e}ncia e Tecnologia, I.P. project Grant No. IN1234CEECINST/00099/2021 and through the FCT projects UIDB/04564/2020, UIDP/04564/2020, and CERN/FIS-PAR/0027/2021, with DOI identifiers 10.54499/UIDB/04564/2020, 10.54499/UIDP/04564/2020, and 10.54499/CERN/FIS-PAR/0027/2021, respectively. The work of F.C. is supported by the European Union's Horizon 2020 research and innovation programme under the Marie Sklodowska-Curie grant agreement No 860881-HIDDeN.

 \end{document}